\documentclass{article}
\usepackage{spconf,amsmath,graphicx,url}
\usepackage{booktabs}
\usepackage{color}
\usepackage{mathtools}
\usepackage[hidelinks]{hyperref}
\usepackage[sort]{cite}

\title{BETTER TOGETHER: DIALOGUE SEPARATION AND VOICE\\
ACTIVITY DETECTION FOR AUDIO PERSONALIZATION IN TV}
\name{Matteo Torcoli and Emanuël A. P. Habets\thanks{$^*$A joint institution of Fraunhofer IIS and Friedrich-Alexander-Universität Erlangen-Nürnberg (FAU).}}
\address{International Audio Laboratories Erlangen$^*$, Am Wolfsmantel 33, 91058 Erlangen, Germany.}

\usepackage[pscoord]{eso-pic}
\newcommand{\placetextbox}[3]{
\setbox0=\hbox{#3}
\AddToShipoutPictureFG*{
\put(\LenToUnit{#1\paperwidth},\LenToUnit{#2\paperheight}){\vtop{{\null}\makebox[0pt][c]{#3}}}%
}%
}%

\begin{document}

\maketitle

\begin{abstract}
In TV services, dialogue level personalization is key to meeting user preferences and needs. When dialogue and background sounds are not separately available from the production stage, Dialogue Separation (DS) can estimate them to enable personalization.
DS was shown to provide clear benefits for the end user. Still, the estimated signals are not perfect, and some leakage can be introduced. This is undesired, especially during passages without dialogue. We propose to combine DS and Voice Activity Detection (VAD), both recently proposed for TV audio. When their combination suggests dialogue inactivity, background components leaking in the dialogue estimate are reassigned to the background estimate. A clear improvement of the audio quality is shown for dialogue-free signals, without performance drops when dialogue is active.
A post-processed VAD estimate with improved detection accuracy is also generated. It is concluded that DS and VAD can improve each other and are better used together.
\end{abstract}
\placetextbox{0.5}{0.08}{\fbox{\parbox{\dimexpr\textwidth-2\fboxsep-2\fboxrule\relax}{\footnotesize \centering Accepted paper. \copyright  2023 IEEE. Personal use of this material is permitted. Permission from IEEE must be obtained for all other uses, in any current or future media, including reprinting/republishing this material for advertising or promotional purposes, creating new collective works, for resale or redistribution to servers or lists, or reuse of any copyrighted component of this work in other works.}}}
\section{Introduction}
\label{sec:intro}

Movies, series, documentaries, news, and TV shows constitute an important part of our daily lives. 
A vital role in the TV audio quality is played by the balance between foreground speech (referred to as dialogue) and background sounds \mbox{\cite{torcoli2019preferred, shirley:2004}}.
Difficulties in following dialogue due to loud background sounds have been a known issue for over three decades~\cite{Mathers:1991}. 
Dialogue Enhancement (DE) allows the user to personalize the audio and adjust the relative dialogue level to suit individual needs and preferences. DE can be provided by Next Generation Audio (NGA), e.g., MPEG-H Audio~\cite{Simon:2019}.  

When dialogue and background are not separately available from the production stage, Dialogue Separation~(DS) can estimate them, given only the final soundtrack as input. 
DE has been shown to reduce listening effort and improve user satisfaction, also when enabled by DS~\cite{torcoli2018adjustment, IBC:2021, westhausen2021reduction, torcoli2022}.
Nevertheless, the separated signals are not ideal, and some distortions can be introduced in the personalized soundtrack, which is especially undesired during passages without dialogue.
One cause for these distortions is background components leaking into the separated dialogue while dialogue is not active.
The audio elements causing leakage often share temporal and spectral characteristics with speech and pose particular challenges for DS.
Modifying or retraining DS to be more aggressive with these elements often results in inferior speech quality.

In this paper, we focus on improving the quality of passages free of target signal (dialogue in our case), which are often disregarded while evaluating source separation algorithms~\cite{schulze2019weakly}.
In particular, a post-processing method for DS is proposed. First, dialogue-free passages are detected based on the synergy between DS and Voice Activity Detection (VAD). When the combination of DS and VAD indicates a dialogue-free passage with high confidence,
the signal components leaking in the dialogue estimate are reassigned to the background estimate.
The reported experiments consider state-of-the-art Deep Neural Networks (DNNs) for DS  \cite{paulus22sampling} and VAD \cite{hung2022large} proposed for TV services.
The evaluation is carried out on an openly available dataset for TV audio~\cite{petermann2022cocktail}.
A clear improvement in the audio quality of the separated signals is shown when no dialogue is present, while minimal differences are observed when dialogue is present.
The envelope of the post-processed dialogue is used to generate a new VAD estimate with detection accuracy higher than both VAD and DS considered individually.

\section{Dialogue Separation (DS)}

DS produces an estimate of the dialogue signal $\hat{d}(n)$ and an estimate of the background $\hat{b}(n)$ given the input mixture
\begin{equation}
x(n) = d(n) + b(n),
\end{equation}
where $d(n)$ and $b(n)$ are the corresponding unobservable reference signals. It is required that $\hat{d}(n)+\hat{b}(n)  =x(n)$. The estimates can be used to enable personalization. The balance between dialogue and background can be controlled by the user by means of remixing gains $g(n)$, e.g.:
\begin{equation}
y(n) = \hat{d}(n) + g(n)\hat{b}(n).
\end{equation}
The personalized output $y(n)$ is typically normalized to obtain a constant integrated loudness. 
The final user is not meant to listen to $\hat{d}(n)$ and $\hat{b}(n)$ separately. 
In contrast to speech enhancement for telecommunications, some amount of background is desired in TV audio, given its narrative importance \cite{ward2019}. 
Hence, some degree of remixing is applied, and $g(n)$ is limited to avoid full suppression of the components \cite{torcoli2021controlling}. The gains $g(n)$ can be constant or smoothly change over time (dynamic). Dynamic gains attenuate the estimated background only when the estimated dialogue exceeds certain thresholds~\cite{IBC:2021}. Using dynamic gains can also help in reducing perceivable degradations when the estimated dialogue is not active.

Previous DS solutions leverage characteristics specific to dialogue in TV productions, e.g., the fact that speech is usually amplitude-panned in a stereo scene \cite{master2020}, typically in the virtual center \cite{geiger2015}, or that it is a direct component \cite{craciun2015}, or a combination of these characteristics \cite{paulus:2019}. 
A more general approach is proposed in \cite{uhle:2008}, where feature extraction is followed by a shallow neural network. More recent works use DNNs for DS, e.g., \cite{paulus22sampling, westhausen2021reduction, petermann2022cocktail, petermann2022tackling}. 

In \cite{petermann2022tackling, master2020, geiger2015,uhle:2008}, VAD algorithms are also discussed along with DS systems. In these works, VAD is used as conditional input to DS or it is used to apply soft-gating to the DS estimates, or to process the mixture $x(n)$ only when dialogue is detected and apply broadband attenuation otherwise. Three aspects differentiate this paper from previous works. First, we consider state-of-the-art DS and VAD specifically proposed for TV audio. Second, we propose a novel combination of DS and VAD, where the VAD output is not used directly, but consensus is sought between DS and VAD to: i)~detect dialogue activity with higher accuracy, ii)~post-process the separated outputs based on combined activity information, and iii)~generate a refined VAD. Third, the experimental evaluation gives particular attention to dialogue-free passages.

As a baseline for DS, we consider the fully convolutional network described in \cite{paulus22sampling}, operating in the STFT-domain on stereo input audio at a sampling rate of 48\,kHz.
As an update to \cite{paulus22sampling}, the network is trained using more training material, i.e., almost 31 hours of real-world stereo TV productions internally cleaned as in~\cite{IBC:2021}. The training data is augmented, similarly to \cite{paulus22sampling}, by selecting a random mixing gain $\in [-12, +6]$\,dB, a random overall gain $\in [-6, +6]$\,dB, a random offset in the beginning (max.\,10\,ms), and 33\% chance of being downmixed to mono.

\section{Voice Activity Detection (VAD)}

VAD takes $x(n)$ as input and outputs a continuous dialogue presence probability $\hat{p}(n)$ $\in [0, 1]$ that can be thresholded to obtain a binary detection signal $\hat{v}(n)$ $\in \{0, 1\}$, where as threshold we used $0.5$.
Early VAD algorithms are based on various types of extracted features, e.g., energy levels \cite{loizou2007speech}, the Itakura LPC spectral distance \cite{rabiner1977voiced}, or spectral flux \cite{geiger2015}. More recent approaches adopt DNNs \cite{hung2022large, braun2021training, jo2021self, petermann2022tackling}.

As a baseline, we adopt the convolutional recurrent neural network proposed in \cite{hung2022large}, trained to perform VAD along with music activity detection. The network is trained on around 1600\,h of professional audio for TV shows.
For this training material, noisy labels are derived from different sources, e.g., subtitles, cue sheets, or pre-trained model predictions. This detector is referred to as TV Speech and Music (TVSM). We use the speech detection of the so-called \textit{pseudo} TVSM model as provided by the authors\footnote{Pre-trained model available at \url{https://github.com/biboamy/TVSM-dataset/}}.

\section{Signal Component Reassignment (SCR)}
\label{sec:method}
The goal of the proposed SCR is to reassign leaking background components from the dialogue estimate to the background estimate. The SCR is applied to the outputs of the DS system.
SCR can be implemented using a broadband reassignment gain $r_\textrm{S}(n) \in [0, 1]$:
\begin{equation}
 \hat{d}_\textrm{R}(n) = (1-r_\textrm{S}(n)) \hat{d}(n),
\end{equation}
\begin{equation}
    \hat{b}_\textrm{R}(n) = r_\textrm{S}(n) \hat{d}(n) + \hat{b}(n),
\end{equation}
where $\hat{d}_\textrm{R}(n)$ and $\hat{b}_\textrm{R}(n)$ are the estimated dialogue and background signals after SCR. 
The  reassignment gain $r_\textrm{S}(n)$ can be obtained in different ways. Three approaches are considered that make use of the VAD in different ways, named \mbox{\textit{DS+VAD-d}}, \mbox{\textit{DS+VAD-v}}, and \mbox{\textit{DS+VAD-p}}.
A number of parameters are introduced to describe these approaches (see Table\,\ref{tab:impl}). The parameters were empirically determined using a few development items.

\begin{table}[t]
	\caption{Parameter values used in our implementation.
	\label{tab:impl}}
\centering
\begin{footnotesize}
\begin{tabular}{c c c c c}
        \toprule
$P_{\textrm{in}, \textrm{min}}$ & $P_{\textrm{in}, \textrm{max}}$  &  $P_{\textrm{out}, \textrm{min}}$ & $P_{\textrm{out}, \textrm{max}}$ & $W_{\textrm{len}}$  \\
        \midrule
$0.3$ & $0.7$ & $0 $& $2$ & $600$\,ms  \\
        \bottomrule
\end{tabular}
\begin{tabular}{c c c c c c}
        \toprule
$T_{\textrm{rel}}$ & $T_{\textrm{abs}, z}$  & $T_{\textrm{abs}, v}$ & $C_\textrm{f}$ & $W_{\textrm{fill}}$  & $T_r$ \\
        \midrule
$-20$\,dB & $-45$\,dB &   $-40$\,dB & $6.9\times10^{-5}$ & $500$\,ms & 0.2\\
        \bottomrule
        
\end{tabular}
\end{footnotesize}
\end{table}

The most elementary approach, referred to as \mbox{\textit{DS+VAD-d}}, is to directly gate the DS estimate by means of $\hat{v}(n)$, i.e.,
\begin{equation}
r_\textrm{S}(n) = 1 - \hat{v}(n).
\end{equation}
Gating can introduce abrupt changes in the output signal. To improve the perceived quality, smoothing can be applied.

In the second considered approach, named \mbox{\textit{DS+VAD-v}}, $r(n)$ is first obtained as  \mbox{$r(n) = 1 - \hat{v}(n)$} and then smoothed to obtain $r_\textrm{S}(n)$.
The smoothing is implemented using a forward-backward first-order infinite impulse response (IIR) filter with $C_\textrm{f}$ as feedback coefficient, and $r_\textrm{S}(n)$ is set to $0$ whenever $r_\textrm{S}(n)<T_r$.

\begin{figure}[tb]
  \centerline{\includegraphics[width=\columnwidth]{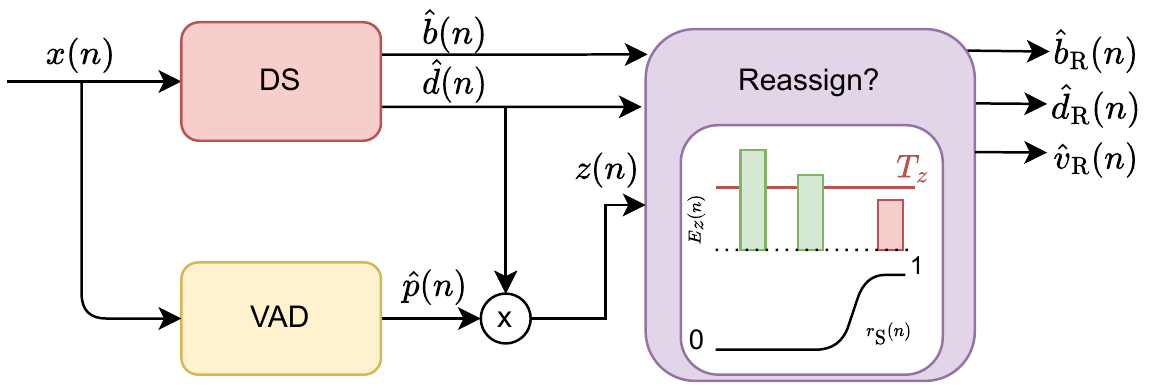}}
\caption{\textit{DS+VAD-p}: DS and VAD are combined to detect dialogue activity with higher accuracy. When no dialogue is present, components leaking in the dialogue estimate are reassigned to the background, and a refined VAD is generated.}
\label{fig:method}
\end{figure}

We consider a third approach to carry out SCR, referred to as \textit{DS+VAD-p}. This makes use of the continuous presence probability $\hat{p}(n)$. As depicted in Figure \ref{fig:method}, the outputs of DS and VAD are combined to generate a control signal
\begin{equation}
z(n) = \hat{p}(n) \hat{d}(n).
\label{eq:gating}
\end{equation}
The envelope of $z(n)$,
i.e., $E_z(n)$, 
can be interpreted as a combined estimate of dialogue activity. We compute the envelope as root-mean-square envelope with a sliding window of length $W_{\textrm{len}}$. 
Hence, a reassignment decision $r(n)$ is taken:
\begin{equation}
r(n)=
\begin{dcases}
    1,		& E_z(n) < T_z\\
    0,            & \text{otherwise}.
\end{dcases}
\end{equation}
As threshold, we use $T_z = \max(T_{\textrm{rel}} \overline{E}_z,\,  T_{\textrm{abs}, z})$, where $\overline{E}_z$ is the envelope mean.
The same smoothing already described is applied to $r(n)$ so to obtain $r_\textrm{S}(n)$.
\textit{DS+VAD-p} can be tuned to weigh the contributions of DS and VAD to the reassignment decision. To do so, we propose to clip $\hat{p}(n)$ between $P_{\textrm{in}, \textrm{min}}$ and $P_{\textrm{in}, \textrm{max}}$ and to linearly rescale it to $P_{\textrm{out}, \textrm{min}}$ and $P_{\textrm{out}, \textrm{max}}$ before applying Eq.\,(\ref{eq:gating}).

Finally, $\hat{d}_\textrm{R}(n)$ is used to generate a refined VAD signal $\hat{v}_\textrm{R}(n)$. To this end, the envelope of $\hat{d}_\textrm{R}(n)$ is thresholded by $T_v = \max(T_{\textrm{rel}} \overline{E}_{\hat{d_\textrm{R}}},\,  T_{\textrm{abs}, v})$. 
Gaps of size up to $W_{\textrm{fill}}$ between detected activities (\mbox{$\hat{v}_\textrm{R}(n)=1$}) are filled with ones to avoid too frequent changes in the refined VAD output.

\begin{figure}[t]
\centerline{\includegraphics[width=\columnwidth]{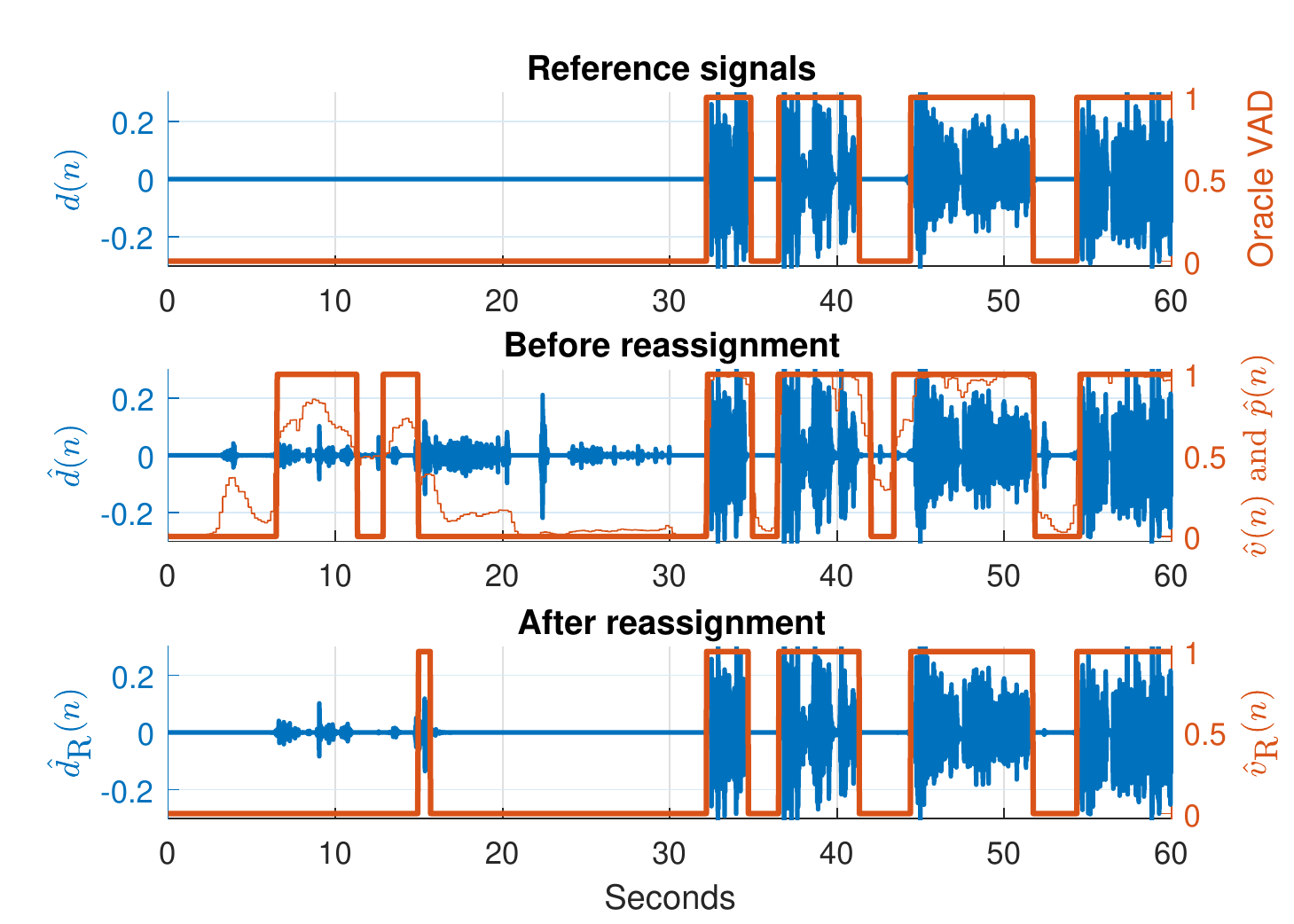}}
\caption{Upper plot: reference dialogue signal $d(n)$ and oracle VAD. Middle plot: dialogue estimate $\hat{d}(n)$ and VAD outputs $\hat{p}(n)$ and $\hat{v}(n)$. 
Lower plot: post-processed signals.
}
\label{fig:idea}
\end{figure}

Figure \ref{fig:idea} shows an example applying \textit{DS+VAD-p}\footnote{Audio examples are also available at: {\url{https://www.audiolabs-erlangen.de/resources/2022-DS-TVSM-SCR}}}. A mixture $x(n)$ is considered in which no dialogue is active for the first 30\,s. 
Music and effects are active for the full duration of the example. 
The reference $d(n)$ and $v(n)$ are depicted in the upper subplot. 
As shown by $\hat{d}(n)$ (middle plot), DS is able to accurately estimate the dialogue parts.
However, leaking components can be observed, especially in the first 30\,s. The VAD output also misclassifies some of these components. SCR is applied to the passages in which $\hat{p}(n)$ is low and $\hat{d}(n)$  exhibits low energy. 
The lower plot shows that $\hat{d}_\textrm{R}(n)$ has been cleaned from a relevant amount of leaking components, while leaving the dialogue parts unaltered. The refined VAD $\hat{v}_\textrm{R}(n)$ improves accuracy and resolves some misclassifications observed in $\hat{v}(n)$. 

\section{Experiments}
\label{sec:experiments}

\subsection{Evaluated Methods}

We evaluate the three SCR approaches described in Sec.\,\ref{sec:method}, along with the baseline DS and VAD without post-processing.  
As an additional baseline, an SCR method that does not make use of an external VAD is considered. This method is referred to as \mbox{\textit{DS+threshold}} and consists of the thresholding and smoothing operations carried out for \textit{DS+VAD-p}, but without the use of $\hat{p}(n)$, i.e., Eq.\,(\ref{eq:gating}) becomes  $z(n) = \hat{d}(n)$.
Finally, as the upper performance bound, the oracle VAD $v(n)$ is considered, where $v(n)$ is obtained by thresholding the reference dialogue $d(n)$, i.e., the envelope of $d(n)$ is computed, thresholded, and gap-filled using $W_{\textrm{len}}$, $T_{\textrm{rel}}$, $T_{\textrm{abs}, v}$, and $W_{\textrm{fill}}$.
The SCR method named \textit{DS+oracleVAD} uses $v(n)$ for computing the reassignment gains as $r_\textrm{S}(n) = 1 - v(n)$.

\subsection{Test Data and Scenarios}

The Divide and Remaster (DnR) test set~\cite{petermann2022cocktail} was used, resampled to 48\,kHz. This dataset was recently released to support the development of methods for separating and remixing TV audio
and is openly available\footnote{\url{https://doi.org/10.5281/zenodo.6949108}}. 
The test set consists of 973 mono mixtures of 60\,s. For each mixture, three separate mono signals are provided: dialogue, music, and effects. We mixed music and effects without altering their ratio, forming MUSFX, i.e., dialogue-free signals.
The MUSFX signals contain heterogeneous types of music and effects, and possibly also background speech (e.g., unintelligible chattering or chanting), making the task for DS and VAD particularly challenging.
For the input signals, we consider two test scenarios:
\begin{enumerate}
\item MIX: The MUSFX signals were mixed with the accompanying dialogue signals.
In the original test set, most of the loudness differences between dialogue and MUSFX are limited to a small range centered around 4\,LU (Q1\,$=2.9$, Q3\,$=5.1$\,LU), which is not representative of the full range encountered in TV material~\cite{torcoli2019preferred}. Therefore, we report evaluation metrics for the full unmodified test set as well as for the first hour, for which the mix of dialogue and MUSFX is done with five different SNR levels from $-5$ to $+15$\,dB with steps of $5$\,dB. This MIX scenario was evaluated by comparing estimated and reference dialogue.

\item MUSFX: Only the MUSFX signals are considered, without dialogue. In the evaluation, the estimated background is compared with the reference MUSFX.
\end{enumerate}

\section{Results}

Performance is measured using: the 2f-model without boundary detection \cite{torcoli2021controlling}, SI-SDR \cite{leroux2019}, wideband PESQ \cite{P862revised}, and VAD accuracy, i.e., $ (TP + TN ) / N$, reported as a percentage, where $N$ is the number of frames, and $TP$ and $TN$ are true positives and true negatives, respectively. Results are reported in Tables \ref{tab:results_MIX} and \ref{tab:results_MUSFX} for the MIX and MUSFX scenarios, respectively.
As in \cite{petermann2022tackling}, all outputs are normalized to have the same integrated loudness as their corresponding ground truth.

\begin{table}[t]
\setlength{\tabcolsep}{3.5pt}
	\caption{Results for MIX on the full original DnR test set and on the first hour with augmented SNR. Best scores in bold (disregarding the oracle condition). \mbox{\textit{DS+VAD-p}} shows slight improvement on all metrics with respect to the baselines.
	\label{tab:results_MIX}}
\centering
\begin{footnotesize}
\begin{tabular}{l c c c c}
        \toprule
 & 2f & SI-SDR& PESQ & VAD acc.\,(\%)\\ 
        \midrule
Input           &  $ \,\,\,6.3 \,|\, 10.2 $   & $1.0 \,|\, 5.0 $ & $\,1.24 \,|\, 1.52$ & --\\
DS (baseline)& $ 21.4 \,|\, 25.7 $ & $\,\,\,8.0 \,|\, 10.5$ &  $\,1.43 \,|\, 1.79$ & --\\
VAD (baseline) & -- & -- & -- & $\,91.5 \,|\, 91.2$  \\
DS+threshold & $22.0 \,|\, 27.2$           & $\,\,\,8.1 \,|\, 10.6 $   &   $\,1.44 \,|\, 1.83$    & $\,93.6 \,|\, 95.0$   \\
DS+VAD-d & $23.4 \,|\, 28.6 $ & $\,\,\,8.2 \,|\, 10.6$ & $\mathbf{1.46} \,|\, 1.83$   & $\mathbf{95.7} \,|\, 95.7$ \\
DS+VAD-v & $ \mathbf{23.6} \,|\, \mathbf{28.7} $ & $\,\,\,8.2 \,|\, 10.3$ & $\,\mathbf{1.46} \,|\, \mathbf{1.84}$   & $ \mathbf{95.7} \,|\, 95.7$ \\
DS+VAD-p & $\,23.4 \,|\, \mathbf{28.7} $ & $\,\,\,\mathbf{8.4} \,|\,  \mathbf{10.8} $ & $\,\mathbf{1.46} \,|\, \mathbf{1.84} $  & $\,\mathbf{95.7} \,|\,\mathbf{96.1} $\\
DS+oracleVAD & $ 25.3 \,|\, 30.1 $ & $\,\,\,8.8 \,|\, 11.1 $ & $\,1.48 \,|\, 1.84 $ & $\,98.9 \,|\, 98.5$ \\
        \bottomrule
\end{tabular}
\end{footnotesize}
\end{table}

For the MIX scenario (Table\,\ref{tab:results_MIX}), the baseline DS clearly improves the performance with respect to the input mixtures, although it was trained on stereo signals and it is here tested on mono inputs. 
\textit{DS+VAD-v} is the only method that slightly reduces the performance compared to the baseline, namely the SI-SDR on the augmented first hour of test material.
All other SCR methods do not deteriorate the baseline performance. 
\textit{DS+VAD-p} exhibits slightly better performance than \textit{DS+VAD-d} and \textit{DS+threshold}, and only marginally worse than \textit{DS+oracleVAD}. \textit{DS+VAD-p} is preferred over \mbox{\textit{DS+VAD-d}} also because the first one applies smoothed SCR, while the latter does not. This difference is hardly captured by averaged metrics, but it can correspond to clearly superior perceived quality locally. 
It is also worth noting that although the baseline VAD accuracy is lower than \textit{DS+threshold}, combining DS and VAD via SCR exhibits superior VAD accuracy than both DS and VAD individually.

The MUSFX scenario is where the benefits of the proposed SCR methods are expected.
Table\,\ref{tab:results_MUSFX} shows that DS degrades the objective quality with respect to the input. Dialogue is not present, so the input equals the reference signal in this scenario.
The proposed DS+VAD methods successfully improve the quality by a significant margin with respect to the baselines, i.e., more than $10$ MUSHRA points estimated by 2f, more than $12$\,dB in terms of SI-SDR, more than $1.2$ MOS points estimated by PESQ, and $2.6$ to $3.9$ \% better VAD accuracy. In this test scenario, \mbox{\textit{DS+VAD-v}} shows better performance than \textit{DS+VAD-p}. Still, \textit{DS+VAD-p} is preferred due to its advantages for the MIX scenario.
\textit{DS+threshold} shows only marginal improvements with respect to DS, reflecting the benefits of integrating an independent VAD. As already observed for MIX, combining DS and VAD shows superior accuracy compared to either DS and VAD separately.

\begin{table}[t]
	\caption{Results for MUSFX on the full original DnR test set. Best scores in bold. \mbox{\textit{DS+VAD-p}} improves the DS baseline more than $10$ MUSHRA points as estimated by 2f, more than $12$\,dB in terms of SI-SDR, and more than $1.2$ PESQ points.
	\label{tab:results_MUSFX}}
\centering
\begin{footnotesize}
\begin{tabular}{l c c c c}
        \toprule
 & 2f & SI-SDR& PESQ & VAD acc.\,(\%)\\ 
        \midrule
Input           &  $ 100 $   & +Inf & $4.6$ & --\\
DS (baseline)& $ 52.3 $ & $ 12.5 $ &  $ 2.67 $ & --\\
VAD (baseline) & -- & -- & -- & $ 91.9$  \\
DS+threshold & $ 54.6 $           & $ 14.0$   &   $ 3.28 $    & $ 87.3 $   \\
DS+VAD-d & $62.4$ & $ 27.3$ & $3.95$   & $ 95.5$ \\
DS+VAD-v & $\mathbf{66.5} $ & $ \mathbf{28.1}$ & $ \mathbf{4.04}$   & $ \mathbf{95.8}$ \\
DS+VAD-p & $ 62.7 $ & $ 24.7 $ & $ 3.93 $  & $ 94.5 $\\
DS+oracleVAD & $ 100 $ & +Inf & $ 4.6 $ & $ 100$ \\
        \bottomrule
\end{tabular}
\end{footnotesize}
\end{table}

\section{Conclusions}
With audio personalization for TV in focus, a rule-based post-processing method for DS is proposed. State-of-the-art DNNs for DS and VAD are combined, 
dialogue-free passages are detected, and leaking components are reassigned. The proposed method improves not only the DS output quality but also the VAD accuracy, suggesting that DS and VAD can complement each other and are better used together.

Future work should consider training a post-processing network taking as inputs the DS and VAD outputs, and developing DNNs capable of DS with less background leaking without degrading the perceived speech quality.

\section{Acknowledgment}
Sincere thanks go to the Fraunhofer IIS Accessibility Solutions team and to Jouni Paulus for the valuable suggestions.

\bibliographystyle{IEEEbib}
\bibliography{references}

\end{document}